\documentclass[11pt,preprintnumbers,nofootinbib,prd]{revtex4}
\usepackage{graphicx}
\usepackage{amsmath}
\usepackage{epsfig}
\usepackage{mathrsfs}

\newcommand{\be}{\begin{equation}}
\newcommand{\ee}{\end{equation}}
\newcommand{\ba}{\begin{eqnarray}}
\newcommand{\ea}{\end{eqnarray}}

\begin{document}
\input{epsf}
\begin{flushright}
SLAC-PUB-13681
\end{flushright}

\title{Scanning the Fluxless $G_2$ Landscape}
\author{Konstantin Bobkov}
\affiliation{Stanford Linear Accelerator Center,\\ 2575 Sand Hill Road, Menlo Park, CA 94025}

\vspace{0.5cm}

\date{\today}

\vspace{0.3cm}

\begin{abstract}
We show that there exists an exponentially large discretuum of vacua in $G_2$-compactifications of M-theory
without flux. In M-theory-inspired $G_2$-MSSM, quantities relevant for particle physics remain virtually
insensitive to large variations of the vacuum energy across the landscape. The purely non-perturbative 
vacua form a special subset of a more general class of vacua containing fractional Chern-Simons contributions. 
The cosmological constant can be dynamically neutralized via a chain of 
transitions interpolated by fractional gauge instantons describing spontaneous nucleation of M2-brane domain 
walls. Each transition is generically accompanied by a gauge 
symmetry breaking in some sector of the theory. In particular, the visible sector 
GUT symmetry breaking can likewise be triggered by a spontaneous nucleation of an M2-brane.

\end{abstract}
\maketitle
\newpage
\vspace{-1.2cm} \tableofcontents

\section{Introduction}\label{intro}
The existence of an exponentially large, and possibly infinite, landscape of vacua in string theory
seems beyond doubt \cite{Susskind:2003kw}. The key ingredient in generating such a landscape is the presence of discrete fluxes,
coming from both open and closed string sectors, needed to freeze some or all compactification moduli \cite{Dasgupta:1999ss},
\cite{Giddings:2001yu}, \cite{Acharya:2002kv}, \cite{DeWolfe:2005uu}, \cite{Kachru:2003aw}.
At present, the landscape paradigm presents the only known solution \cite{Bousso:2000xa}, \cite{DeSimone:2008bq} 
to the cosmological constant problem \cite{Weinberg:1988cp}. Since string theory has no continuous adjustable parameters, the only
possibility to implement the fine tuning of the cosmological constant is via a discrete scan over the integer flux data. 
Moreover, such a scan is fully dynamical because transitions between different flux vacua can be naturally 
described by a non-perturbative mechanism involving membrane nucleation \cite{Bousso:2000xa}, \cite{Feng:2000if}, \cite{Brown:1987dd}.

If we embrace the claim that the tiny value of the cosmological constant is attainable only in those string compactifications
which result in an exponentially large landscape of densely spaced discrete vacua, we are led to a puzzle with regard to
the recently constructed fluxless de Sitter vacua \cite{Acharya:2007rc}, \cite{Acharya:2008hi} in $G_2$-compactifications 
of M-theory. Naively, it appears as if the absence of fluxes in such models renders the mechanism of discrete tuning 
of the vacuum energy impossible. Therefore, the purpose of this note is to explicitly address the question of discrete tuning of 
the cosmological constant in this novel class of M-theory vacua. 

One of the most attractive properties of the fluxless vacua is the natural
explanation of the gauge hierarchy between the Planck and electroweak scales. This feature is entirely due
to the fact that in the absence of tree-level flux contributions, the entire superpotential is purely non-perturbative.
Strong gauge dynamics in the hidden sectors naturally generates a hierarchically small supersymmetry
breaking scale in the visible sector via the mechanism of dimensional transmutation \cite{Witten:1981nf}. 
At the same time, the superpotential generically depends on all compactification 
moduli which in M-theory correspond to volumes of three-cycles. 
In \cite{Acharya:2007rc}, \cite{Acharya:2008hi} it was demonstrated that in addition to
solving the hierarchy problem, all the moduli can be explicitly stabilized in the regime where 
supergravity approximation can be trusted.

In the series of papers \cite{Acharya:2008hi}, \cite{Acharya:2008zi}, \cite{Acharya:2008bk}, \cite{Kumar:2008vs}, \cite{Kane:2009kv},
phenomenology and cosmology of these vacua was studied in detail. In particular, soft terms
in the supersymmetry breaking lagrangian were explicitly computed and the spectrum of
superpartners at the electroweak scale was studied. The most generic feature of the $G_2$-MSSM
sparticle spectrum is the prediction of relatively light gauginos $m_{1/2}\sim {\cal O}(100){\rm GeV}$
and heavy squarks and sleptons $m_0\sim{\cal O}(1-10){\rm TeV}$\footnote {In \cite{Acharya:2008hi}, the so-called approximately
sequestered limit was also considered where all superpartners, including scalars, are 
relatively light with masses ${\cal O}(100-2000)\,{\rm GeV}$.}.
In \cite{Acharya:2008bk} moduli couplings to visible sector matter were computed and 
it was demonstrated that the predominantly non-thermal dark matter production mechanism
results in the relic density being very close to the observed value. Furthermore, a natural 
solution to the cosmological gravitino problem was found, mainly due to a unique hierarchy in
the moduli mass spectrum \cite{Acharya:2008bk}.

In a more recent development, local M-theory models with a realistic three-generation MSSM-like spectrum 
have been successfully constructed \cite{Pantev:2009de}, \cite{Bourjaily:2009vf}. Such models contain a natural solution to the doublet-triplet splitting
problem, the $\mu$-problem and have a mechanism for generating a realistic texture of the Yukawa couplings.

Here we demonstrate that the fluxless vacua, constructed in \cite{Acharya:2007rc}, \cite{Acharya:2008hi}, generate an
exponentially large landscape where discrete Wilson line and topological data play the role of discrete
quanta. We show explicitly that by dialing only a few integer parameters, the leading contribution to the cosmological constant 
can be tuned to $V_0\approx {\cal O}(10^{-3})\times m_{3/2}^2m_{pl}^2\approx 10^{-31}\times m_{pl}^4$ and that further fine-tuning 
of the vacuum energy has virtually no effect on the predictions for particle physics in the M-theory-inspired $G_2$-MSSM.
We suggest that tuning the cosmological constant to the observed value may be achieved by scanning over the exponentially large 
discretuum.

We also consider a natural extension of the purely non-perturbative vacua by including tree-level contributions from the 
fractional Chern-Simons invariants. We argue that the corresponding de Sitter vacua automatically end up in the
domain of small volume and appear to be highly metastable. We explain how the tree-level contributions can be reduced 
dynamically via spontaneous nucleation of M2-branes interpolated by fractional gauge instantons. In particular, when 
crossing the bubble wall separating different vacua, the Page charge jumps by a rational number, while the vacuum energy
density is typically shifted by $\delta V_0 \sim M_{GUT}^4$. As a result, after a chain of such transitions
the net Chern-Simons term may be completely neutralized while the universe ends up in one of the exponentially 
many purely non-perturbative vacua \cite{Acharya:2007rc}, \cite{Acharya:2008hi}.
Furthermore, due to the change in the discrete Wilson line data across the domain wall, each bubble nucleation is typically accompanied 
by a gauge symmetry breaking in some sector of the theory, including the visible sector.
In other words, the visible sector GUT symmetry breaking via discrete Wilson line may play an important role in neutralizing the large tree-level 
value of the cosmological constant and may, in fact, be the final step in the above chain of transitions.

Thus, we present a dynamical mechanism which interpolates between two very different classes of vacua - 
the highly metastable GUT-scale vacua, which may realize some form of inflation, 
and the fluxless vacua \cite{Acharya:2007rc}, \cite{Acharya:2008hi}, which exhibit a large hierarchy between 
the Planck and electroweak scales stabilized by the low-scale supersymmetry.

This paper is complimentary to  \cite{Acharya:2007rc}, \cite{Acharya:2008hi}, \cite{Acharya:2008zi}, \cite{Acharya:2008bk}, \cite{Kumar:2008vs}, \cite{Kane:2009kv}, and is organized as follows. In section \ref {sec2} we compute threshold corrections to the gauge kinetic functions
of some simple hidden sectors in $G_2$ compactifications of M-theory. In particular, we consider $SU(N)$ gauge groups
that arise from the breaking $SO(2N)\rightarrow SU(N)\times U(1)$ and rely on the methods of \cite{Friedmann:2002ty}
to derive an explicit expression for the one-loop thresholds due the massive Kaluza-Klein modes living
on the hidden sector three-cycles. In section \ref{sec3} we present a quick overview of the fluxless
vacua constructed earlier in \cite{Acharya:2007rc}, but taking into account the modifications in \cite{Acharya:2008hi}.
Then, using the results derived in section \ref{sec2}, in section \ref{sec4} we explicitly demonstrate how the 
discrete tuning of the cosmological constant can be implemented by scanning over the discrete 
Wilson line and topological data. At the same time, we observe that obtaining robust predictions in
the $G_2$-MSSM requires very minimal tuning of the vacuum energy, i.e. very large variations of the cosmological constant
have virtually no effect on the corresponding particle physics. We estimate the number of non-perturbative vacua 
to be exponentially large and suggest that the observed value of the cosmological constant may be realized in a large 
subset of such vacua. Finally, in section \ref{sec5} we propose a mechanism to describe the dynamics of transitions towards 
the purely non-perturbative vacua from the more general class of vacua possessing a non-zero 
Chern-Simons contribution at tree-level via spontaneous nucleation of M2-branes.
We also find a strong hint that the visible sector GUT symmetry breaking could have been triggered 
by such a membrane nucleation and was simultaneously accompanied by a large decrease in the vacuum energy.

\section {Threshold corrections}\label{sec2}

The number one question we have to address is where the exponentially large discretuum can possibly come 
from when the fluxes are turned off? At first sight, it seems as though apart from the discrete group-theoretic 
quantities such as quadratic Casimirs contained inside the gaugino condensates, if one regards them as tunable,
there appears to be nothing else to tune. However, upon closer examination one discovers that threshold 
corrections to the gauge kinetic functions of the condensing gauge theories may
contain a number of independent discrete dials that come in a form of discrete Wilson line
and topological data \cite{Friedmann:2002ty}. Therefore, motivated by this observation, obtaining explicit 
expressions for such thresholds will be the main task of this section.

In our analysis we will closely follow the techniques of \cite{Friedmann:2002ty} where 
one-loop Kaluza-Klein threshold corrections to the visible sector gauge couplings were 
first computed in the context of $SU(5)$ SUSY GUT model\footnote{A more general result for the threshold 
corrections in $SU(N)\times SU(M)\times U(1)$ theories without charged matter obtained from the breaking 
$SU(N+M)\rightarrow SU(N)\times SU(M)\times U(1)$ was presented in Tamar Friedmann's Ph.D. thesis \cite{Friedmann:2002ty}.} 
arising from $G_2$ compactifications of M-theory. In such constructions the GUT gauge group is broken to the Standard Model 
by a discrete Wilson line. As argued in \cite{Friedmann:2002ty}, these threshold corrections can be expressed in terms 
of certain topological invariants and have a remarkable property of preserving the unification of the 
MSSM gauge couplings\footnote{This statement only holds for $SU(5)$  \cite{Friedmann:2002ty}. However, this property
is in contrast to the recently constructed F-theory GUT models \cite{Beasley:2008dc}, where the $SU(5)$ is broken by the 
internal hypercharge flux, which induces splitting of the gauge couplings at the string scale \cite{Blumenhagen:2008aw}.}. 

In M-theory compactified on a singular $G_2$ holonomy manifold $X$, non-Abelian gauge fields live in
${\bf R}^{3,1}\times {\cal Q}$ where ${\cal Q}\in X$ is an associative three-cycle along which there is an orbifold
singularity of A-D-E type \cite{bsa}, \cite{Acharya:2000gb}, \cite{Atiyah:2000zz}, 
\cite{Acharya:2001hq}, \cite{Atiyah:2001qf}. In addition, there may exist point-like conical singularities
at isolated points $p_i\in X$ such that some $p_j\in\cal Q$. At $p_j$ the rank of the singularity 
gets enhanced and depending on the type of enhancement one obtains charged chiral matter in some
irreducible representation of the gauge group supported along $\cal Q$ 
\cite{Atiyah:2001qf}, \cite{Witten:2001uq}, \cite{Acharya:2001gy}, \cite{Acharya:2004qe}. For instance, if
the singular locus $\cal Q$ supports $SO(10)$, there may exits an isolated point 
$p\in {\cal Q}$ where the singularity is enhanced to either $SO(12)$ or $E_6$ in which case one obtains charged
chiral matter in either $10$ or $16$ of $SO(10)$ localized at $p$.

In what follows, we compute one-loop threshold corrections to the gauge couplings of $SU(N)$ hidden 
sector gauge theories coming from the massive Kaluza-Klein modes where the co-dimension four singular 
locus ${\cal Q}\in X$ is a lens space. In particular, we consider $SU(N)$ gauge theories originating 
from breaking
\be
SO(2N)\rightarrow SU(N)\times U(1)\,,
\ee
by a discrete Wilson line. As we will see below, when $SU(N)$ gauge theories are obtained 
in this way, the Kaluza-Klein states can give rise to substantial threshold corrections to the 
$SU(N)$ gauge kinetic function. In turn, these thresholds slightly modify the non-perturbative
terms of the superpotential and as a result, directly affect a certain parameter 
of the theory which controls the leading order contribution to the cosmological constant.
Furthermore, if there exists at least one codimension seven conical singularity at point 
$p\in {\cal Q}$, such that at $p$ the singularity gets enhanced to $SO(2N+2)$, 
a chiral matter multiplet transforming in $2N$ of $SO(2N)$ will arise at $p$ and after the symmetry 
breaking will automatically give us the $N$ and $\bar N$ of $SU(N)$ which will then form 
the effective meson field $\phi$ and generate the matter F-term $F_{\phi}$ necessary for the existence
of de Sitter vacua \cite{Acharya:2007rc}, \cite{Acharya:2008hi}.

The authors of \cite{Friedmann:2002ty} argued that in $G_2$ compactifications of 
M-theory, threshold corrections from the Kaluza-Klein modes living on a supersymmetric three-cycle $\cal Q$
can be explicitly computed even without knowing the $G_2$ metric. More specifically, such corrections come 
in a form of linear combinations of Ray-Singer analytic torsions \cite{Ray} which are topological 
invariants of ${\cal Q}$. We should also stress that the three-cycles considered in \cite{Friedmann:2002ty}
are rigid, i.e. $b_1({\cal Q})=b_2({\cal Q})=0$, and have a non-trivial fundamental group.

Following the notation of \cite{Friedmann:2002ty}, the threshold correction to the Wilsonian
holomorphic gauge coupling can be expressed in 
terms of Ray-Singer torsion classes ${\cal T}_i$ as
\be\label{tc}
{\cal S}^{\prime}_a=2\sum_i{\cal T}_i {\rm Tr}_{{\cal R}_i}Q^2_a\,,
\ee
where $Q_a$ is a generator of the a-th gauge group of the hidden sector and ${\cal R}_i$ is the i-th
representation.

Let us consider a singular seven-dimensional manifold $X$ with $G_2$ holonomy which in some
local neighborhood can be represented as a product ${\bf R}^4/D_N\times{S^3}/{Z_q}$. 
Here $D_N$ is the binary dihedral group acting on ${\bf R}^4$ while $Z_q$ is a cyclic group 
acting on the three-sphere. 
Moduli spaces of such compactifications were studied in detail in \cite{Friedmann:2002ct} and \cite{Ita:2002ws}.
As a result, we obtain an $SO(2N)$ codimension four singularity whose three-dimensional locus
${\cal Q}=S^3/Z_q$ is a lens space. In this case $b_1({\cal Q})=b_2({\cal Q})=0$.

M-theory compactifications on such singular manifolds give rise to 
$SO(2N)$ non-Abelian gauge fields living on ${\bf R}^{3,1}\times {\cal Q}$.
The three-sphere can be described by embedding into the covering space ${\bf C}^2$ as 
\be
|z_1|^2+|z_2|^2=1\,,
\ee
where $z_i$ are complex coordinates. The action of the cyclic group $Z_q$ on the coordinates 
of the covering space is given by
\be
\gamma: z_i\rightarrow e^{i\frac{2\pi w}q}z_i\,,\,\,\,\,\,z_i\in {\bf C}^2\,,
\ee
such that integers $w$ are relatively prime to $q$. Then, the quotient ${\cal Q}=S^3/Z_q$
defines a lens space.

Note that the fundamental group $\pi_1({\cal Q})=Z_q$ and therefore $\gamma \in \pi_1({\cal Q})$ gives the 
action of the fundamental group on the coordinates of the covering space and
results in the following action on the harmonic k-forms $\phi_k$ on ${\cal Q}$:
\be
\gamma\cdot\phi_k\equiv\phi_k(\gamma\cdot z_i)=\omega(\gamma)\phi_k(z_i)\,,\,\,\,\gamma \in \pi_1({\cal Q})\,,\,\,\,z_i\in {\bf C}^2\,,
\ee
where $\omega(\gamma)$ is a one-dimensional complex representation of $\pi_1({\cal Q})$.
When a representation $\omega$ is non-trivial the corresponding Ray-Singer analytic torsion is given by \cite{Ray}
\be\label{nontrivtor}
{\cal T}\left(S^3/Z_q,\omega \right)=\ln|\omega-1|\ln|\omega^{-1}-1|\,.
\ee
On the other hand, using the result of Friedmann and Witten, the analytic torsion
for the trivial representation is \cite{Friedmann:2002ty}
\be\label{trivtor}
{\cal T}_{\cal O}=-\ln q\,.
\ee

Since $\pi_1({\cal Q})=Z_q$, the three-cycle ${\cal Q}$ can support discrete 
Wilson lines which can break the $SO(2N)$ gauge symmetry.

We would like to choose a discrete Wilson line such that we have an embedding of $\pi_1({\cal Q})$
into $SO(2N)$, so that we can determine $\omega(\gamma)$ for the harmonic k-forms $\phi_k$'s transforming in
various $SU(N)$ representations. 
This embedding can be accomplished if we choose the following basis for $2N$ of $SO(2N)$:
\ba
&&y_j=x_{2j-1}+ix_{2j}\,,\,\,\,\,j=\overline{1,N}\,\\
&&\bar y_k=x_{2k-1}-ix_{2k}\,,\,\,\,\,k=\overline{1,N}\,.\nonumber
\ea
This choice of basis also allows us to explicitly split the $2N$ of $SO(2N)$
into $N$ and $\bar N$ of $SU(N)$.

Before we construct the discrete Wilson line, let us recall the decompositions of the fundamental, 
spinor and adjoint representations
of $SO(2N)$ under $SU(N)\times U(1)$:
\ba
&&2N=N_{2m}\oplus \bar N_{-2m}\,,\\
&&{N\,\,{\rm is}\,\,{\rm even}:}\,\,\,2^{N-1}=[0]_{Nm}\oplus[2]_{(N-4)m}\oplus[4]_{(N-8)m}\,.\,.\,.\,\oplus[N]_{-Nm}\,,\nonumber\\
&&{N\,\,{\rm is}\,\,{\rm odd}:}\,\,\,2^{N-1}=[0]_{Nm}\oplus[2]_{(N-4)m}\oplus[4]_{(N-8)m}\,.\,.\,.\,\oplus[N-1]_{(-N+2)m}\,,\nonumber\\
&&Adj_{SO(2N)}=(Adj_{SU(N)})_0\oplus(1)_0\oplus\left(\overline{\frac{N(N-1)}{2}}\right)_{-4m}\oplus\left({\frac{N(N-1)}{2}}\right)_{4m}\,,\nonumber
\ea
where in the decomposition of the spinor representation $[n]$ denotes an $n$-index antisymmetric tensor 
representation of $SU(N)$.

Next we would like to determine the smallest possible value of the overall $U(1)$ charge normalization constant $m$ 
such that all $SU(N)$ multiplets have integer values of the $U(1)$ charge. It turns out that
this depends on whether $N$ is even or odd.
Namely, if $N$ is odd, the decomposition of the spinor representation of $SO(2N)$ always contains an $SU(N)$ 
multplet whose absolute value of the $U(1)$ charge is the smallest, i.e. the corresponding $U(1)$ charge
is either $m$ or $-m$. Thus, for odd $N$ the smallest value of 
$m$ for which all $SU(N)$ multiplets have integer values of the $U(1)$ charge is $m=1$. On the other hand, when $N$ is
even the multiplets whose absolute value of the $U(1)$ charge is the smallest have $U(1)$ charge $2m$ and $-2m$.
Therefore, when $N$ is even, the obvious choice for the minimal value of $m$ is $m=\frac 12$. To summarize, the minimal
value of $m$ such that all $U(1)$ charges have integer values is
\ba\label{mchoice}
&&{N\,{\rm is}\,\,{\rm even}:}\,\,\,\,\,\,\,\,\,m=\frac 12\,,\\
\nonumber\\
&&{N\,{\rm is}\,\,{\rm odd}:}\,\,\,\,\,\,\,\,\,m=1\,.\nonumber
\ea

We can now construct a discrete Wilson line which breaks $SO(2N)\rightarrow SU(N)\times U(1)$
and at the same time represents a generator of $\pi_1({\cal Q})$ embedded into $SO(2N)$
\be
U_{\gamma}=\exp\left(2\pi i(w/q){\rm diag}(2m,\,.\,.\,.\,2m,\,-2m,\,.\,.\,.\,-2m)\right)\,,
\ee
where $Y={\rm diag}(2m,\,.\,.\,.\,2m,\,-2m,\,.\,.\,.\,-2m)$ is the generator of the $U(1)$ charge.
In this case, the action of the fundamental group on a harmonic k-form $\phi_k$ transforming in
a particular irreducible representation of $SU(N)\times U(1)$ whose $U(1)$ charge is $Y_{\phi}$ is given by
\be
\phi_k(\gamma\cdot z_i)=e^{\frac{2\pi iw}q Y_{\phi}}\phi_k(z_i)\,.
\ee
In particular
\ba\label{trprop}
&&\phi_k(\gamma\cdot z_i)=\phi_k(z_i)\,,\,\,\,\,\,\,\,\,\,\,\,\,\,\,\,\,\,\,\,\,\,\,\,\,\,\,\,\,\,\,\,\phi_k(z_i)\in (Adj_{SU(N)})_0\oplus(1)_0\\\nonumber\\
&&\phi_k(\gamma\cdot z_i)=e^{\frac{8\pi imw}q}\phi_k(z_i)\,,\,\,\,\,\,\,\,\,\,\,\,\,\,\phi_k(z_i)\in \left({\frac{N(N-1)}{2}}\right)_{4m}\nonumber\\\nonumber\\
&&\phi_k(\gamma\cdot z_i)=e^{-\frac{8\pi imw}q}\phi_k(z_i)\,,\,\,\,\,\,\,\,\,\,\phi_k(z_i)\in \left(\overline{\frac{N(N-1)}{2}}\right)_{-4m},\nonumber\\\nonumber
\ea
where integer $w$ is such that $4mw$ is not divisible by $q$.

We are now ready to compute the threshold corrections to the gauge kinetic function.
The usual normalization convention for the $SU(N)$ generators $Q_a$ gives
\ba\label{casim}
&&{\cal R}_i=Adj_{SU(N)}:\,\,\,\,\,\,\,\,\,\,\,\,\,{\rm Tr}_{{\cal R}_i}Q^2_a=N\,,\\
\nonumber\\
&&{\cal R}_i=\frac{N(N-1)}2:\,\,\,\,\,\,\,\,\,\,{\rm Tr}_{{\cal R}_i}Q^2_a=\frac{N-2}2\,,\nonumber\\
\nonumber\\
&&{\cal R}_i=N:\,\,\,\,\,\,\,\,\,\,\,\,\,\,\,\,\,\,\,\,\,\,\,\,\,\,\,\,\,\,\,\,\,{\rm Tr}_{{\cal R}_i}Q^2_a=\frac 12\,.\nonumber
\ea

From (\ref{tc}) and (\ref{casim}) we can then compute the threshold correction to the gauge coupling
of $SU(N)$ due to the massive Kaluza-Klein harmonics of the gauge fields.
Using the transformation properties (\ref{trprop}) of the fields arising from decomposition of the adjoint of 
$SO(2N)$ in terms of $SU(N)\times U(1)$ multiplets we obtain:
\be\label{t1}
{\cal S}^{\prime}_{SU(N)}=2N{\cal T}_{\cal O}+(N-2){\cal T}\left(S^3/Z_q,e^{-\frac{8\pi imw}q} \right)+(N-2){\cal T}\left(S^3/Z_q,e^{\frac{8\pi imw}q} \right)\,.
\ee
Using (\ref{nontrivtor}) we obtain
\be\label{t2}
{\cal T}\left(S^3/Z_q,e^{-\frac{8\pi imw}q} \right)={\cal T}\left(S^3/Z_q,e^{\frac{8\pi imw}q} \right)=\ln\left(4\sin^2(4\pi mw/q)\right)\,.
\ee
Combining (\ref{t1}), (\ref{t2}) and (\ref{trivtor}) we obtain
\be\label{t3}
{\cal S}^{\prime}_{SU(N)}=-2N\ln q+2(N-2)\ln\left(4\sin^2(4\pi mw/q)\right)\,.
\ee

In order to construct the second hidden sector we assume that the same manifold $X$ has another codimension four
singularity whose singular locus ${\cal Q}^{\prime}$ is a disjoint copy of ${\cal Q}$ but where a
different binary dihedral group $D_M$ acts on ${\bf R}^4$ so that locally $X$ looks like 
${{\bf R}^4}/{D_M}\times{\cal Q}^{\prime}$. If ${\cal Q}^{\prime}$ contains no codimension seven conical singularities,
the gauge theory one obtains in the second hidden sector upon compactifying M-theory on 
$X$ is a ``pure glue'' $SO(2M)$ SYM gauge theory living in ${\bf R}^{3,1}\times {\cal Q}^{\prime}$.

Again, we can choose a discrete Wilson line to break $SO(2M)\rightarrow SU(M)\times U(1)$ and
use the results we have derived for the first hidden sector. 
Thus, the threshold corrections to the gauge kinetic function of $SU(M)$ from the KK modes
of the gauge fields are given by
\be\label{t4}
{\cal S}^{\prime}_{SU(M)}=-2M\ln q+2(M-2)\ln\left(4\sin^2(4\pi nl/q)\right)\,,
\ee
where $n=1/2$ if $M$ is even and $n=1$ if $M$ is odd. Integer $l$ is such that $4nl$ is not divisible by $q$.

It is interesting to note that in M-theory, the Ray-Singer torsion also appears in the computations
of the zero-mode determinants in the membrane-instanton superpotentials \cite{Harvey:1999as}\footnote{In some literature, the 
Ray-Singer torsions are defined as $e^{\cal T}$.}. In fact, its contribution comes in the same form, i.e. as a 
multiplicative factor $e^{\cal T}$, as in the case of gaugino condensates after one takes into account 
the Kaluza-Klein threshold corrections to the gauge kinetic function of the condensing gauge theory.

\section{Fluxless de Sitter vacua}\label{sec3}

In this section we give a quick overview of the moduli stabilization mechanism developed in
\cite{Acharya:2007rc}, \cite{Acharya:2008hi}.
One of the most distinct features of the construction is the absence of fluxes
which are typically used in string compactifications to stabilize some or all moduli 
\cite{Dasgupta:1999ss}, \cite{Giddings:2001yu}, \cite{Acharya:2002kv}, \cite{DeWolfe:2005uu}.
In M-theory, by choosing to work in the zero flux sector one is forced to use superpotentials that
are purely non-perturbative. The absence of perturbative contributions is due to the
Peccei-Quinn symmetry of the axions, which can only be broken by non-perturbative
effects.
In principle, manifold $X$ may contain many rigid supersymmetric cycles supporting
gaugino condensates and M2-brane instantons. The most general form of the superpotential
is given by \cite{Affleck:1983mk}
\be\label{super}
W=\sum_k A_k\phi_k^{a_k}e^{ib_kf_k}\,,
\ee
where for the $k$-th sector: $A_k$ is a moduli-independent normalization constant,
$\phi_k$ are charged chiral matter fields, $f_k$ is the moduli-dependent gauge kinetic 
function, $b_k$ is either $2\pi$ for instantons or $2\pi/c_k$ for gaugino condensates, 
where $c_k$ is a casimir of the gauge group.
   
It turns out that in order to fix all the moduli in the regime where supergravity
approximation is valid, it is enough to assume two hidden sectors. Such a truncation
is certainly valid as long as the remaining non-perturbative terms are either zero
because the corresponding $<\phi_k>=0$ or exponentially suppressed relative to the 
two leading ones. In the next section we demonstrate that there exists a regime
in the parameter space of the model when the truncated non-perturbative terms 
are exponentially suppressed relative to the leading contributions.

Hence, we consider a scenario where the dominant contributions to the superpotential come
from two hidden sectors with gauge groups $SU(P+N_f)$ and $SU(M)$ where the first hidden sector
gauge theory is a super QCD with $N_f=1$ flavor of quarks $Q$ and $\tilde Q$ transforming in 
a complex (conjugate) representation of $SU(P+1)$ and the second hidden sector with the gauge group $SU(M)$
is a ``pure glue'' super Yang-Mills theory.

The non-perturbative effective superpotential generated by the strong gauge dynamics in the hidden sectors is given by
\be\label{supot}
W=A_1\phi^{-2/P}e^{i\frac{2\pi}{P}f}+A_2e^{i\frac{2\pi}{M}f}\,.
\ee
In the above expression we explicitly assumed that the associative cycles supporting both hidden sectors
are in proportional homology classes which results in the gauge kinetic function being given by essentially 
the same integer combination
of the moduli $z_i$ for both hidden sectors
\be\label{gaugefunction}
f=\sum_{i=1}^{N}N_iz_i\,.
\ee
Here $N=b_3(X)$ and ${\rm Im}(f)=V_{\cal Q}$ is the volume of the corresponding associative cycle.
The complexified moduli $z_i$ are
\be
z_i=t_i+is_i\,,
\ee
where $t_i$ are the axions and $s_i$ are the geometric moduli (volumes of three-cycles) 
describing deformations of the metric of $X$.
The matter field $\phi$ represents an effective meson degree of freedom defined 
as
\be\label{mesdef}
\phi\equiv \sqrt{2Q\tilde Q}
\ee
 in terms of the chiral matter fields $Q$ and $\tilde Q$.

The normalization constants contain the threshold corrections to the gauge kinetic function
and are given by
\be\label{rat}
A_1=CPe^{-\frac{{\cal S}^{\prime}_{SU(N)}}{2P}}\,,\,\,\,\,\,\,A_2=CMe^{-\frac{{\cal S}^{\prime}_{SU(M)}}{2M}}\,,
\ee
where $C$ is the overall normalization, which may be obtained by dimensionally reducing the seven-dimensional
gaugino interaction term.
In \cite{Acharya:2007rc} it was explained that if one uses
a superpotential of the form (\ref{supot}), de Sitter vacua arise only when $M>P$ (if we
include matter in both hidden sectors dS vacua exist without such condition).
Hence, we will keep this in mind from now on.

The total Kahler potential - moduli plus hidden sector matter \cite{Acharya:2008hi}, is given by
\be\label{Kahler}
K = -3 \ln 4\pi^{1/3} V_X+\frac{\bar\phi\phi}{V_X}\,,
\ee
where the seven-dimensional volume $V_X$ is a homogeneous function of 
the moduli $s_i$ of degree $7/3$ \cite{Beasley:2002db}.

The full expression for the ${\cal N}=1$ $D=4$ supergravity scalar potential is quite complicated and
we choose not to exhibit it here. In our analysis of the moduli stabilization in \cite{Acharya:2008hi} 
we did not rely on a specific form of the seven-dimensional volume $V_X$ as a function of the moduli $s_i$. 
We simply used the fact that the volume $V_X$ is a homogeneous function of $s_i$.
In particular, we were able to convert the problem of determining the moduli vevs at the minimum 
of the scalar potential into a more simple problem of determining the values of certain parameters ${\tilde a}_i$,
satisfying a constraint
\be
\sum_{i=1}^N{\tilde a}_i=\frac 73\,.
\ee

At the minimum of the potential ${\tilde a}_i$ are constants and can be found by solving
a system of $N$ transcendental equations
\be\label{eqforatilde}
\frac{{\tilde a}_i}{3N_i}\hat K_i|_{_{s_i=\frac{{\tilde a}_i}{N_i}}}
+{\tilde a}_i\,=0\,,\,\,{\rm no\,\,sum\,\,over\,\,i}\,,
\ee
where $\hat K_i$ is the derivative of the bulk Kahler potential $\hat K = -3 \ln 4\pi^{1/3} V_X$
with respect to $s_i$. As can be seen from the formulas that follow, for the moduli to be stabilized at
positive values, solution of (\ref{eqforatilde}) must be such that
\be\label{acon}
{\tilde a}_i>0\,,\,\,\forall{i}\,,\,{i=\overline{1,N}}\,.
\ee
In \cite{Acharya:2008hi} we checked for rather generic choices of parameters when the volume 
of $X$ is given by
\be
V_X=\sum_k c_k\prod_{i=1}^Ns_i^{a_i^k}\,,\,\,\,\,{\rm where}\,\,\,\forall k\,:\,\sum_{i=1}^Na_i^k=\frac 73\,.
\ee
as well as by more complicated homogeneous functions of degree $7/3$, 
that such solutions indeed exist and are quite common.
Once the values of ${\tilde a}_i$ are found, the moduli vacuum expectation values (vevs) at the
minimum are given by
\be\label{ans2}
s_i=\frac{{\tilde a}_i}{N_i}\frac 37V_{\cal Q}\,.
\ee
where $V_{\cal Q}$ is the volume of the associative cycle supporting the hidden sectors.
It is stabilized at the value
\be\label{volhid}
V_{\cal Q}\approx\frac{PM}{2\pi(M-P)}\ln\left(\frac{MA_1\phi_0^{-2/P}}{PA_2}\right)\,.
\ee

It is convenient to define a parameter
\be\label{peffdef}
P_{eff}\equiv P\ln\left(\frac{MA_1\phi_0^{-2/P}}{PA_2}\right)\,.
\ee
As we will see below, when the vacuum energy is tuned to be small, the value of $P_{eff}$ becomes 
fixed at $P_{eff}\sim {\cal O}(60)$ - which is quite large. Hence, $1/P_{eff}$ is a natural small 
parameter of our model. In the leading order, the moduli vevs are given by
\be\label{modvev}
s_i=\frac{{\tilde a}_i}{N_i}\frac{3MP_{eff}}{14\pi(M-P)}\,.
\ee

We note that since ${\tilde a}_i>0$, $N_i>0$ and  $M>P$, we need $P_{eff}>0$ so that there exists a
local minimum with $s_i>0$.

Using the moduli vevs and dropping the terms of order ${\cal O}(1/P_{eff}^2)$,
the scalar potential at the minimum as a function of the meson field is \cite{Acharya:2008hi}
\be\label{vacen}
V_0=\left[\left(\frac 2{M-P}+\frac{\phi_0^2}{V_X}\right)^2+\frac {14}{P_{eff}}\left(1-\frac 2{3(M-P)}\right)\left(\frac 2{M-P}+\frac{\phi_0^2}{V_X}\right)-3\frac{\phi_0^2}{V_X}\right]\frac{V_X}{\phi_0^2}m_{3/2}^2\,.
\ee
The polynomial in the square brackets in (\ref{vacen}) is quadratic with respect to the canonically normalized meson vev squared $\phi_{c}^2\equiv\phi_0^2/V_X$ with the coefficient of the $(\phi_0^2/V_X)^2$ monomial being positive (+1) and therefore, the minimum $V_0$ is positive when the corresponding discriminant is negative. Tuning the cosmological constant to zero is then equivalent to setting the discriminant of the above polynomial to zero, which boils down to a simple condition
\be\label{peffanalyt}
P_{eff}=\frac{14(3(M-P)-2)}{3(3(M-P)-2\sqrt{6(M-P)})}\,.
\ee 
Note that $P_{eff}$ defined in (\ref{peffdef}) is actually dependent on $\phi_0$. However, 
the approximation $P_{eff}\approx const$ turned out to be self-consistent since 
$P_{eff}$ is fairly large and  $\phi_0^{-2/P}$ appears under the logarithm.
From (\ref{peffanalyt}) we see immediately that
\be
P_{eff}>0\,\Rightarrow\,M-P\geq 3\,.
\ee
Minimizing (\ref{vacen}) with respect to $\phi_{c}^2$ and imposing the constraint $V_0=0$, we obtain the meson 
vev at the minimum in the leading order
\be\label{phian}
\phi_{c}^2=\frac{\phi_0^2}{V_X}\approx\frac 2{M-P}+\frac 7{P_{eff}}\left(1-\frac 2{3(M-P)}\right)\,.
\ee
Hence, if we tune the tree-level vacuum energy for the minimum value $M-P=3$ we obtain from (\ref{peffanalyt}) and (\ref{phian})
\be
P_{eff}\approx 63.5\,,\,\,\,\,\frac{\phi_0^2}{V_X}\approx 0.75\,.
\ee

Although the value $\phi_{c}^2\approx 0.75$ appears to be somewhat large, one should recall that
the definition of the meson field (\ref{mesdef}) contains a factor of two.
Hence, along the D-flat direction, the bilinears of the chiral matter fields which appear in the 
Kahler potential are $<Q_{c}^{\dagger}Q_{c}>=<{\tilde Q}_{c}^{\dagger}{\tilde Q}_{c}>  \approx 0.37$, which is
a bit more comforting since we explicitly neglected possible higher order terms in the Kahler 
potential for hidden sector matter.

By minimizing the scalar potential numerically we find that for the minimum value $M-P=3$, the tuning of the cosmological 
constant by varying the constants $A_1$ and $A_2$ inside the superpotential results in fixing the value of $P_{eff}$ at
\be\label{CCcond}
P_{eff}\approx 61.648\,,
\ee
while the canonically normalized meson vev squared is stabilized at
\be\label{mvev}
\phi_{c}^2=\frac{\phi_0^2}{V_X}\approx 0.746\,,
\ee
thus confirming the analytical results above. 

As was demonstrated in \cite{Acharya:2007rc}, \cite{Acharya:2008hi},
the minimal allowed value $M-P=3$ is the only case when the moduli vevs can be large enough
so that sufficient control in the supergravity regime can be achieved. 
Hence, we keep our focus on this minimal choice.

Yet, far more importantly, this case automatically leads to a prediction of TeV-scale
supersymmetry by constraining $m_{3/2}\sim{\cal O}(10)\,{\rm TeV}$ and 
$m_{1/2}\sim{\cal O}(100)\,{\rm GeV}$ \cite{Acharya:2008hi}. 

Furthermore, explicit computations have revealed that despite the presence of a very large number of microscopic
parameters whose values are kept arbitrary, the soft susy-breaking terms have a surprisingly simple structure.
For example, it turns out that the soft terms are independent of the integer parameters 
$N_i$ ($N_i^{\rm vis}$) appearing in the gauge kinetic functions of the hidden (visible) 
sectors\footnote{The anomaly-mediated contributions to the gaugino masses do depend on the volume
of the visible sector associative cycle which is given by a linear combination
$\alpha_{GUT}^{-1}=\sum_{i=1}^N N_i^{\rm vis}s_i$.}. Nor do they depend on the total number of moduli.

A property dubbed ``Kahler independence'' was recently discovered in \cite{Acharya:2008hi}. Namely, it was shown that
the dependence of the soft terms on the form of the moduli Kahler potential 
appears only as dependence on the seven-dimensional volume $V_X$ but not on the manifold-specific 
details of $V_X$ as a function of the moduli\footnote{In fact, with the exception of the gravitino mass, 
the volume $V_X$ always appears in combination with the meson field vev as $\frac{\phi_0^2}{V_X}$, which
is completely fixed by (\ref{mvev}).}. From the phenomenological standpoint $V_X$ is a single parameter, 
fixed once the moduli are fixed. Thus, apart from very few parameters which appear in
the superpotential, the vast number of microscopic details of the compactification
manifold remain completely hidden.
All the above features make the predictions of the $G_2$-MSSM universally valid for all 
$G_2$ compactifications compatible with the above construction.

\section{The discretuum}\label{sec4}
In this section we would like to demonstrate that discrete tuning of the cosmological constant
is possible in the context of fluxless compactifications of \cite{Acharya:2007rc}, \cite{Acharya:2008hi},
described in the previous section. Here, the role of discrete dials is played by both topological 
and Wilson line data.

In fact, it turns out that the problem of tuning the vacuum energy can be naturally broken into two categories.
First, it is clear that the largest contribution to the vacuum energy arises from the two leading 
non-perturbative terms in the superpotential which fix the moduli. As we saw in the previous section,
in order to cancel this energy one needs to tune parameter $P_{eff}\approx 62$. In this section we 
explicitly show that this can be done. We refer to this tuning as {\it coarse} tuning.

The second question is whether the cosmological constant can be discretely tuned to the observed value, 
once all other contributions are taken into account. This tuning is referred to as {\it fine} tuning.
To address this problem we outline a possible mechanism which involves a scan over discrete data in 
the truncated part of the superpotential. However, a definite answer to this question requires a 
more comprehensive statistical study which we leave for future work.

\subsection{Coarse tuning}
As a first step we would like to point out that the question of {\it fine} 
tuning of the cosmological constant can be effectively separated from the problem of obtaining robust 
predictions for particle physics.
This property can be seen from the following simple observation.
Note that the value of $P_{eff}$ not only controls the leading contribution to the
vacuum energy but also explicitly enters into the analytic expressions for the soft breaking terms
\cite{Acharya:2008hi} in a form of a small parameter $1/P_{eff}$.\footnote{ 
Note that the value of $P_{eff}$ required to tune the leading contribution to the vacuum energy 
$P_{eff}\approx 62$ is quite large.}
As evident from the corresponding formulas \cite{Acharya:2008hi}, small variations in the 
range $61\leq P_{eff}\leq 62$ hardly affect the values of the soft breaking terms.
For example, the tree-level gaugino to gravitino mass ratio, while being the most sensitive to the variations 
of $P_{eff}$, is only changed by about $1.6\%$, as can be inferred from the explicit formula
\be\label{gm}
\frac{m_{1/2}^{tree}}{m_{3/2}}\approx -\frac{1}{P_{eff}}\left(1+\frac{2V_X}{(M-P)\phi_0^2}+{\cal O}\left(\frac 1{P_{eff}}\right)\right)\,.
\ee
Quite dramatically, such small variations in $P_{eff}$ simultaneously result in vastly different values of the vacuum energy
\be
61\leq P_{eff}\leq 62\,\,\,\,\Rightarrow\,\,\,\,-\left(m_{3/2}m_{pl}\right)^2\times 10^{-3}\lesssim V_0\lesssim+\left(m_{3/2}m_{pl}\right)^2\times 10^{-3}\,.
\ee
Therefore, once we demonstrate that we can {\it coarsely} tune $P_{eff}$ to its approximate value, 
the {\it remaining} cosmological constant problem becomes essentially decoupled from most of particle physics.
Hence, the question of {\it coarse} tuning is what we would like to address first.
Recall the definition of $P_{eff}$ given in (\ref{peffdef})
\be
P_{eff}\equiv P\ln\left(\frac{MA_1\phi_0^{-2/P}}{PA_2}\right)\,,
\ee
where we consider $P=N-1\approx N\sim O(10)$. Since $P_{eff}$ turns out to be large we can use in our evaluation
of $P_{eff}$ an approximate expression
\be\label{t5}
P_{eff}\approx P\ln\left(\frac{MA_1}{PA_2}\right)=P\left(\frac {{\cal S}^{\prime}_{SU(M)}}{2M}
-\frac{{\cal S}^{\prime}_{SU(N)}}{2P}\right)\,,
\ee
where in the second equality we used (\ref{rat}). From the above expression
we see that in order to determine $P_{eff}$ we need to use explicit expressions
for the threshold corrections to the tree-level gauge kinetic functions in both hidden sectors.
After substituting (\ref{t3}) and (\ref{t4})
into (\ref{t5}) we obtain
\be\label{pe}
P_{eff}\approx P\,\ln\left(\frac {\left(4\sin^2(4\pi nl/q)\right)^{\frac{M-2}M}}{\left(4\sin^2(4\pi mw/q)\right)^{\frac{P-1}P}}\right)\,.
\ee

\begin{table}[t!]
\label{peffective}
\begin{tabular}{||c||c|c|c|c|c|c|c||}
\hline\hline $P_{eff}$ \rule{0pt}{3.0ex}\rule[-1.5ex]{0pt}{0pt} &
$P$ & $M$ & $q$ & $l$ & $w$ & $n$ & $m$ \\ \hline
\,\,\,61.3\,\,\, & \,\,\,10\,\,\, & \,\,\,13\,\,\, & \,\,\,99\,\,\, & \,\,\,12\,\,\, & \,\,\,25\,\,\, & 1 & 1 \\
\,\,\,61.9\,\,\, & \,\,\,20\,\,\, & \,\,\,23\,\,\, & \,\,\,17\,\,\, & \,\,\,6\,\,\, & \,\,\,4\,\,\, & 1 & 1 \\
\,\,\,64.2\,\,\, & \,\,\,27\,\,\, & \,\,\,30\,\,\, & \,\,\,11\,\,\, & \,\,\,3\,\,\, & \,\,\,5\,\,\, & \,\,\,1/2\,\,\, & \,\,\,1/2\,\,\, \\ \hline\hline
\end{tabular}%
\caption{Values of $P_{eff}$ obtained from (\ref{pe}) for three different sets of integer parameters $P,\,M,\,q,\,w,\,l$. 
For a given $G_2$ manifold the values of $P$, $M$, and $q$ are fixed whereas $l$ and $w$ can be chosen freely.}
\end{table}

Table I demonstrates that by dialing the integers in (\ref{pe}) we can tune $P_{eff}$ very close to the desired value. 
Therefore, the {\it coarse} discrete tuning of the vacuum energy can be achieved.
Including the factor of $\phi^{-2/P}$ into (\ref{pe}) gives a few percent correction. It is clear from (\ref{pe})
and the entries in the table that in order to get a large value of $P_{eff}$ one has a tradeoff between
the values of $P$ and $q$. In particular, for $P\lesssim 10$ the value of $q$ has to be very large, i.e. $q\sim{\cal O}(100)$, 
whereas if $P$ is in the range $20\lesssim P\lesssim 30$, required values of $q$ become much more reasonable.
Generically we expect that a typical $G_2$ compactification contains gauge groups with smaller ranks,
i.e. $SU(N)$ with $N<5$, and therefore values $P,\,M > 20$ are probably not typical. 
It is still unclear if there exists an upper bound on the ranks of non-Abelian gauge groups
in $G_2$ compactifications of M-theory\footnote {In F-theory, compact $CY_4$ examples with gauge groups $SO(176)$,
$SO(128)$, $SO(80)$, etc. certainly exist \cite{Candelas:1997pq}.}. Therefore, it is reasonable to expect
that the large values for $P$ and $M$ may eventually be implemented in explicit $G_2$ compactifications. 
Furthermore, one should keep in mind that the expression in (\ref{pe})
was obtained for a rather specific choice of the hidden sector three-cycle, i.e. the lens space $S^3/Z_q$, and a particular way 
the hidden sector $SU(N)$ was obtained, namely by breaking $SO(2N)\rightarrow SU(N)\times U(1)$. It is quite plausible
that obtaining the value $P_{eff}\approx 62$ by discrete dialing is easier for more general setups. For instance,
it would be very interesting to compute $P_{eff}$ for the cases when one or both hidden sectors originate from breaking an 
exceptional group such as $E_8$.

Next, we would like to show that the ``double condensate'' regime with only two exponentials
in the superpotential can be made parametrically self-consistent.
Recall that the full non-perturbative superpotential (\ref{super}) may contain
a large number of terms. At the same time, from Table I it is now clear that in order to cancel
the leading contribution to the vacuum energy, the two dominant gaugino condensates
which stabilize the moduli must come from gauge groups with fairly large ranks.
More importantly, it turns out that once the value of $P_{eff}$ is fixed,
apart from the normalization constant, the two dominant superpotential terms are also fixed, 
independent of the values of $P$ and $M$! For example, the contribution of the second condensate at the minimum is
\be
\delta W\sim e^{-\frac{2\pi}MV_{\cal Q}}\,.
\ee 
Using the volume of $V_{\cal Q}$ from (\ref{volhid}), for $M-P=3$ and $P_{eff}\approx 60$ we obtain
\be\label{est}
V_{\cal Q}\approx 3M\,,\,\,\,\,\Rightarrow\,\,\,\,\,\delta W\sim e^{-6\pi}\,.
\ee
The fact that the dominant contributions remain fixed can be effectively used to suppress the rest of the
non-perturbative terms as follows. Let us consider another rigid three-cycle ${\cal Y}\in X$ which can support
either an $SU(N)$ gaugino condensate or an instanton. The stabilized volume of such cycle is given by
an integer combination of the moduli, different from (\ref{gaugefunction})
\be
V_{\cal Y}=\sum_{i=1}^NN_i^{\prime}s_i=\frac 37V_{\cal Q}\sum_{i=1}^N\frac{N_i^{\prime}{\tilde a}_i}{N_i}
\approx M\frac 97\sum_{i=1}^N\frac{N_i^{\prime}{\tilde a}_i}{N_i}=M\times{\cal O}(1)\,,
\ee
where we used (\ref{ans2}) and (\ref{est}). The corresponding contribution to the superpotential is proportional to
\be\label{concy}
\delta W_{\cal Y}\sim e^{-\frac{2\pi}NV_{\cal Y}}\sim e^{-2\pi\frac MN\times{\cal O}(1)}\,,
\ee
which can be made exponentially small when $M>>N$, compared to the dominant 
contribution (\ref{est}), which remains fixed. In this regime corrections to the moduli vevs from the remaining 
non-perturbative terms can be safely neglected and the predictions of the model for the soft terms are fairly robust.
Any unwanted large-rank gauge groups can be broken to a product of subgroups with smaller ranks by choosing discrete Wilson lines appropriately 
if the corresponding associative cycles have a non-trivial fundamental group.

Note that if $P$ and $M$ were smaller, it would be harder to justify why only two non-perturbative terms were included
in the superpotential. Therefore, the large values of $P$ and $M$ necessary to obtain $P_{eff}\approx 62$
can be regarded as a virtue, which almost ensures that the ``double condensate'' approximation
is self-consistent.

Possible loop corrections to the moduli Kahler potential are also parametrically suppressed when $M$ is large
\be
\delta K_{\rm 1-loop}\sim \frac{g_{\cal Y}^2}{16\pi^2}\times{\cal O}(1)=\frac 1{4\pi V_{\cal Y}}\times{\cal O}(1)
\sim \frac 1{4\pi M}\times{\cal O}(1)\,.
\ee

\subsection{Fine tuning}
The problem of {\it fine} tuning the cosmological constant can also be naturally addressed once
all non-perturbative superpotential contributions are included. This idea was briefly discussed in
\cite{Acharya:2008hi} and here we rely on the same argument but provide more specific details. 
A generic $G_2$ manifold $X$ may contain
a number of rigid associative cycles with non-trivial topology, e.g. general Seifert-fibered 
three-manifolds. In this case, similarly to the leading exponentials, each additional 
non-perturbative term inside the superpotential
will depend on the corresponding discrete data of its cycle while contributing an exponentially
suppressed amount to the cosmological constant. 

To estimate the number of non-perturbative vacua for a given $G_2$ compactification we
first extend the analysis to include more general three-cycles. Consider a $G_2$ manifold
which in some local neighborhood topologically looks like a product 
${\bf R}^4/Z_p\times S^3/(Z_q\times Z_r)$ \cite{Friedmann:2002ct}. 
In this case we obtain an 
$SU(p)$ supersymmetric gauge theory supported along a three-cycle ${\cal Q}=S^{3}/(Z_q\times Z_r)$ 
where $\pi_1({\cal Q})=Z_q\times Z_r$. Note that the lens spaces
considered earlier are included in this set and correspond to the cases when either 
$r=1,\,q>1$ or $q=1,\,r>1$. 
Since the three-cycle is rigid $b_1({\cal Q})=b_2({\cal Q})=0$ it may potentially generate
non-perturbative contributions in a form of multiple gaugino condensates when the $SU(p)$
gauge group is broken to a product subgroup by discrete Wilson lines.
For this sector of the theory, the total number of inequivalent discrete Wilson line 
configurations, where $p$, $q$, $r$ are relatively prime, 
is given by \cite{Friedmann:2002ct}
\be\label{ndwl}
N_{p,q,r}=\frac{\left(p+qr-1\right)!}{p!\left(qr\right)!}\,.
\ee
If we allow the possibility that the integers $p,\,q,\,r$ can be as large as ${\cal O}(10)$,
the number of inequivalent non-perturbative vacua coming from scanning over discrete Wilson 
line data for a single three-cycle becomes enormous! Indeed, using (\ref{ndwl}) we find 
\be
N_{10,9,11}\approx 3.9\times 10^{11}\,,\,\,\,N_{14,9,11}\approx 2.4\times 10^{15}\,,\,\,\,N_{27,19,22}\approx 2.9\times 10^{40}
\ee
vacua. By a conservative estimate, a generic $G_2$ manifold is expected to have 
$N_{\rm cycles}\lesssim {\cal O}(10)$ associative cycles supporting non-Abelian gauge groups. 
Therefore, the total number of inequivalent vacua for a given manifold can be estimated as
\be\label{nvac}
{\cal N}_{vac}\sim \prod_{i=1}^{\rm N_{cycles}} N_{p_i,q_i,r_i}\sim 10^{{\cal O}(100)}\,.
\ee

Yet, despite this vast landscape, there exists an exponentially large subset where the values of 
the soft-breaking terms, fixed by the {\it coarse} tuning, remain virtually identical in each vacuum!
Hence, within this subset of vacua, while the two dominant condensates 
responsible for the moduli fixing cancel the largest contribution to the vacuum energy,
the subdominant terms may potentially do the rest of the job in canceling the remaining 
vacuum energy yielding the value of the cosmological constant in the observed range.
Whether such an extreme discrete fine tuning is possible remains to be seen because
even with the existence of an exponentially large landscape of vacua, we still need to 
show that the value of the cosmological constant either varies uniformly over the discretuum
or can be peaked in the vicinity of the observed value.
Thus, a comprehensive statistical study would certainly be welcome.

Finally, we would like to point to an important subtlety, which may arise in the above scenario. 
In addition to the potentially large number of non-perturbative terms, topologically non-trivial 
associative cycles supporting discrete Wilson lines may also generate
constant tree-level contributions to the superpotential in a form of fractional Chern-Simons invariants
 \cite{Acharya:2002kv}, \cite{Nishi:1998}, \cite{Gukov:2003cy}.
The presence of such terms would clearly mean a disaster for low-scale supersymmetry, 
since the tree-level contributions would generically overwhelm
the non-perturbative terms and drive the vacua to a regime of small volume where the
${\cal N}=1$ $D=4$ supergravity analysis becomes unreliable. 
Therefore, in order to preserve the non-perturbative vacua one must ensure that 
the total Chern-Simons contribution from all cycles is zero.
Incidentally, some sectors, including the visible sector, do not give non-perturbative
contributions to the superpotential while at the same time they may generate
fractional Chern-Simons terms at tree-level. Thus, one may have enough
freedom to {\it independently} choose discrete data in those sectors in order to kill
the total fractional Chern-Simons contribution while preserving
the fine tuned value of the vacuum energy.

\section{Dynamical neutralization of the vacuum energy}\label{sec5}

Finally we would like to briefly discuss a possible dynamical mechanism realizing the discrete
tuning of the cosmological constant in the context of fluxless M-theory vacua. Recall that the Bousso-Polchinski
mechanism \cite{Bousso:2000xa}, which implements the old idea of Brown and Teitelboim \cite{Brown:1987dd}, 
relies on the spontaneous membrane nucleation in the presence of background fluxes, a phenomenon 
completely analogous to the electron-positron pair production in the presence of very strong electric 
fields - so called Schwinger pair production. When a membrane pair is nucleated, the background flux is decreased by 
an integer unit, thus reducing the total energy. This process continues until the remaining energy is
too small to nucleate another pair. More importantly, in the mechanism of Bousso and Polchinski \cite{Bousso:2000xa}, 
while there exists an exponentially large
number of extremely closely spaced final states for which the total cosmological constant is in the observational range,
the energy density in the vacuum preceding the jump to a final state can be large enough to excite the inflaton away
from its minimum. Hence, the final membrane nucleation may be followed by a subsequent slow roll inflation thus producing
a high enough density of matter and thereby avoiding the ``empty universe'' problem which existed in the 
original Brown-Teitelboim scenario \cite{Brown:1987dd}. Alternatively, even though the inflaton may have been
sitting at the minimum, the inflaton potential itself may be completely modified
after the final jump and therefore a slow roll toward the new minimum may take place.

Can a similar dynamical mechanism exist in the context of fluxless M-theory vacua? Naively, it seems as though the answer
is no because the Bousso-Polchinski mechanism explicitly relies on having background fluxes which couple to
the membranes. Moreover, in order to evade the ``empty universe'' problem, the energy density released in 
the final jump must be large enough to reheat the universe, which may be problematic for the purely
non-perturbative vacua. In addition, the scalar potential in the vicinity of the minimum is generically bounded by 
the gravitino mass scale $V\lesssim {\cal O}(m_{3/2}^2m_{pl}^2)$, clearly indicating that implementing 
high scale inflation where $H_{inf}\sim \sqrt V/m_{pl}\sim {\cal O}(M_{GUT})$ may be virtually impossible
in such models \cite{Kallosh:2004yh}. These considerations lead us to believe that the purely non-perturbative
vacua with low-scale supersymmetry should be regarded as the final states in a chain of multiple transitions
and that the scalar potential describing the moduli dynamics during inflation is different from the
one giving the TeV-scale superpartners. Remarkably,
we already know what the most natural generalization of the purely non-perturbative vacua is! Recalling
the discussion at the end of the previous section, we know that in a generic compactification, the total tree-level
contribution to the superpotential from the fractional Chern-Simons invariants is non-zero. 
In other words, within the entire set of possible fluxless vacua for a given $G_2$ manifold, the purely 
non-perturbative vacua, where the net Chern-Simons contribution is exactly zero, form a special subset, 
albeit an exponentially large one.

In M-theory, the vacua with a non-zero tree-level contribution to the superpotential should in some very superficial
sense resemble the KKLT-type vacua of Type IIB \cite{Kachru:2003aw}\footnote{Rather, before the F-term uplift,
these vacua are more like the the Heterotic vacua \cite{Gukov:2003cy}.}, where in M-theory the constant Chern-Simons 
term plays the same role as the flux-induced Gukov-Vafa-Witten 
superpotential \cite{Gukov:1999ya}, which is effectively a constant, once the complex structure moduli are frozen. However, 
unlike Type IIB, tuning the net fractional Chern-Simons contribution to an exponentially small value in order
to obtain vacua in the regime where the supergravity analysis can be trusted is much harder to justify\footnote{Note that obtaining configurations where the net Chern-Simons contribution is exactly zero is
completely natural.}. Another key difference from the Type IIB case is the absence of the no-scale property, which
implies that at large volume, the scalar potential tends to zero from above which makes the F-term uplifting
mechanism possible and quite natural for such vacua. 

When the tree-level part is present, the total superpotential has the form
\be
W=W_{\rm tree}+W_{\rm np}\,,
\ee
where $W_{\rm tree}$ is the net Chern-Simons contribution and $W_{\rm np}$ is the non-perturbative part given by (\ref{super}).
To check if de Sitter vacua exist we performed a numerical analysis for a simple toy model with two moduli:
\ba
&&K = -3 \ln 4\pi^{1/3} V_X+\frac{\bar\phi\phi}{V_X}\,,\,\,\,V_X=s_1^{a_1}s_2^{a_2}\,,\,\,\,\,a_1+a_2=\frac 73\\
\nonumber\\
&&W=W_{\rm tree}+A_1\phi^{-2/P}e^{i\frac{2\pi}{P}f}+A_2e^{i\frac{2\pi}{M}f}\,,\,\,\,\,f=N_1z_1+N_2z_2\,.\nonumber
\ea
For the following choice of parameters\footnote{The specific value for the Chern-Simons contribution used here was taken from Table V in \cite{Nishi:1998}.}
\be
W_{\rm tree}=-\frac 1{120}\,,\,\,\,A_1=27\,,\,\,\,A_2=14.85\,,\,\,\,P=27\,,\,\,\,M=30\,,\,\,\,N_1=N_2=1\,,\,\,\,a_1=a_2=\frac 76\,,
\ee
we obtain
\be
s_1=s_2\approx 10.22\,,\,\,\,\,\phi_c^2=\frac{\phi_0^2}{V_X}\approx0.58\,,\,\,\,V_0\approx 1.49\times 10^{-14}\times m_{pl}^4\approx\left( {8.5\times 10^{14}\,{\rm GeV}}\right)^4\,,
\ee
while the axions $t_1$, $t_2$ and $\theta$ satisfy the constrains
\be
\cos\left(\frac{2\pi}{P}(N_1t_1+N_2t_2)-\frac 2P\,\,\theta\right)=1\,,\,\,\,\,\,\cos\left(\frac{2\pi}{M}(N_1t_1+N_2t_2)\right)=-1\,.
\ee
Although the moduli vevs are quite small, one can argue that in this toy example they may be large enough to 
trust the supergravity analysis. We don't have a good way to check this statement but we expect that even when 
quantum corrections are included, de Sitter minima should still exist.

Let us now describe a dynamical mechanism, which can automatically neutralize the tree-level term, 
resulting in one of the purely non-perturbative vacua as a final outcome. 
For a three-cycle ${\cal Q}\in X$ supporting the non-Abelian gauge fields, 
the tree-level superpotential contribution in a form of a fractional Chern-Simons invariant $c_1$ is given by
\be\label{csinv}
\delta W_{tree}^{{\cal Q}}=\omega(A,{\cal Q})=\frac 1{8\pi^2}\int_{{\cal Q}}\left(tr A\wedge dA+\frac 23A\wedge A\wedge A\right)\equiv c_1\,,
\ee
where $A$ is a flat gauge connection. 

There is a subtle but impotant point regarding the tree-level Chern-Simons contributions which we would like to briefly discuss.
Recall that the Chern-Simons invariants (\ref{csinv}) can only be determined up to an integer. This ambiguity comes
from the possibility of performing large gauge transformations on the gauge connection with the gauge instanton
interpolating between different gauge field configurations. As a result, the vacuum energy depends on an integer,
which is not fixed for a given value of $c_1$. This seemingly bizarre dependence of the vacuum energy on the large
gauge transformations has a physical interpretation\footnote{Here we rely on essentially the same arguments 
as in \cite{Beasley:2002db}.}. In particular, it turns out that this integer can be interpreted as
the value of the period $\int_{X} G_7/{4\pi^2}$ of the seven-form $G_7$, which is the dual of the four-form $G_7=*G_4$.
The period of the seven-form $G_7$ is not constant because the form is not closed. 
Indeed, in the presence of co-dimension four singularities supporting non-Abelian gauge fields, 
$G_7$ satisfies the following equation of motion \cite{Acharya:2002kv}
\be
dG_7+\frac 12G_4\wedge G_4+\delta^4_Y\wedge tr F\wedge F=0\,,
\ee
where $\delta^4_Y$ is a four-form delta-function supported on $Y=R^{3,1}\times{\cal Q}$. 
The quantity that does remain constant is the conserved Page charge \cite{Page:1984qv}
\be
P=\frac 1{4\pi^2}\int_X\left(G_7+\frac 12C_3\wedge G_4\right)+c_1\,,
\ee
where $C_3$ is the three-form potential $G_4=d C_3$. 

Let us explicitly break the four-form $G_4$ into two parts
\be
G_4=G_4^0+G_4^X\,,
\ee
where $G_4^0=G_4|_{R^{3,1}}$ and $G_4^X=G_4|_{X}$.
For the types of vacua we are considering here, the internal components of the four-form are zero $G_4^X=0$.
In this case, the seven-form $G_7=*G_4=*G_4^0$ has its legs only in the internal directions, which is exactly
what we need in order to compute the Page charge:
\be\label{pch}
P=c_1+\frac 1{4\pi^2}\int_X *G_4^0\,.
\ee
We can now fix the charge at a particular constant value, which is conserved. Then,
the possibility of adding an integer to the Chern-Simons invariant $c_1$ translates into the possibility
of shifting the value of the four-form flux $G_4^0$ by an integer number of quanta.
In fact, this is precisely consistent with the general form of the flux superpotential in $G_2$ compactifications
when the Page charge contribution to the superpotential is non-zero \cite{Beasley:2002db}, \cite{Acharya:2000ps}.
Unfortunately, the vacua where the tree-level part is greater than one automatically lie in the domain of extremely
small volume where quantum corrections become very important and where the simple framework of ${\cal N}=1$ $D=4$ 
supergravity is most likely inadequate. All we can say here is that those vacua should be
highly metastable since the tunneling amplitude for a gauge instanton is quite
large when the volume of the corresponding associative cycle is very small. These transitions
correspond to spontaneous nucleation of M2-brane domain walls in $R^{3,1}$ with
the Page charge jumping by an integer unit when crossing the wall.

From now on we will concentrate on those vacua where the net Chern-Simons contribution is a rational number
less than one. In four dimensions, two different vacua corresponding to different values of the gauge connection are separated by
a domain wall which is effectively described by a gauge field instanton on ${\cal Q}\times R$, where R is a direction
transverse to the wall \cite{Acharya:2002kv}, \cite{Gukov:2003cy}, \cite{Austin:1990ax}. 
Because gauge field configurations with a non-zero fractional instanton number
carry a non-zero M2-brane charge (\ref{pch}), the domain wall corresponds to an M2-brane whose world-volume
is $R^{2,1}\in {\cal Q}\times R^{3,1}$.

The tension of the domain wall separating the vacua is proportional to the instanton action and 
can be expressed as the difference between the Chern-Simons invariants
\be
T\sim\Delta W\sim\frac 1{8\pi^2}\int_{{\cal Q}\times R} tr F\wedge *F=\frac 1{8\pi^2}\int_{{\cal Q}\times R} tr F\wedge F
=\omega(A,{\cal Q})|_{+\infty}-\omega(A,{\cal Q})|_{-\infty}=\Delta c_1=-c_2\,,
\ee
where $c_2$ is a fractional instanton number, which is equal to the shift of the Page 
charge (\ref{pch}) across the domain wall. 
In particular, when the associative cycle ${\cal Q}$ is a lens space ${\cal Q}=S^3/Z_q$
the Chern-Simons invariants are given by \cite{Nishi:1998}, \cite{Gukov:2003cy}
\be
c_1=-\frac {m^2}{4q}-\frac{\lambda^2}2\,\,\,\,\,\,{\rm mod}\,{\rm\bf Z}\,,
\ee
where
\be
\lambda =0\,,\,\frac 12\,\,,
\ee
and $m$ is an integer defined modulo $2q$, i.e. $m\sim m+2q$.\footnote{The moduli space of such instantons is discussed in \cite{Gukov:2003cy}.} Therefore, during each transition, the net tree-level contribution to the superpotential from the 
fractional Chern-Simons invariants is automatically shifted by a rational number $c_2=k/q$ \cite{Gukov:2003cy}. 
Correspondingly, the vacuum energy undergoes a large jump
\be\label{trans}
\Delta W_{tree}\sim \frac kq\,,\,\,\,\,\Rightarrow\,\,\,
\,\Delta V_0\approx \frac 1{64\pi V_X^3}|\Delta W_{tree}|^2\sim \frac {k^2}{q^2}(10^{17}\,{\rm GeV})^4\,,
\ee
where in the above estimate we chose a relatively small seven-dimensional volume $V_X=10$. 
If we ignore gravity and neglect a possible shift in the volume of the three-cycle 
$\cal Q$ during the transition, the tunneling amplitude for the fractional gauge instanton is given by
\be\label{tunamp}
e^{-S_{E}}\sim e^{-\frac{2\pi kV_{\cal Q}}q}\,,
\ee
where the volume of the three-cycle ${V_{\cal Q}}= \frac{4\pi}{g^2}$. The estimate in (\ref{tunamp}) 
indicates that the tunneling can be quite rapid if the integer $q$ is large enough or when
the corresponding vacua are located in the small volume domain $V_{{\cal Q}}\sim O(1)$, 
which is generically inevitable when the tree-level superpotential term is present.
 
Hence, a sequence of such transitions implemented via spontaneous nucleation of M2-branes 
can rather quickly lead to a complete neutralization of the tree-level contribution to the cosmological constant. 
The final transition in this chain should be such that the net Chern-Simons 
term inside the superpotential is exactly zero so that the universe ends up in one of the purely non-perturbative
vacua realizing low-scale supersymmetry. Recall that for a typical $G_2$ compactification,
the number of such final states, estimated in (\ref{nvac}), is exponentially large and potentially contains
a substantial subset of vacua with the values of the cosmological constant compatible with observation. 
It is no less important that during the final transition, the jump in the energy density is typically also very large.
Therefore, assuming we can find a good inflaton candidate \cite{bw:2009} and invoking the arguments of 
Bousso and Polchinski \cite{Bousso:2000xa}, the state preceding the final vacuum has a large enough energy 
density to excite the inflaton away from its minimum. Thus, the subsequent slow roll high-scale
inflation may proceed in the usual fashion, reheating the universe and avoiding the ``empty universe'' problem. 
Note that since this  mechanism interpolates between a scalar potential possessing GUT-scale energy density, which may
potentially drive inflation, and the purely non-perturbative potential that results in low-scale supersymmetry 
breaking, the problem presented by the Kallosh-Linde bound \cite{Kallosh:2004yh} is automatically circumvented.

Finally, we would like to point out to a very intimate connection between the dynamical neutralization
of the cosmological constant described above and the breaking of gauge symmetries by discrete Wilson lines.
Recall that different values of the Chern-Simons invariants are generated
by the different flat gauge connections corresponding to the discrete Wilson lines
in various sectors of the theory, including the visible sector. Thus, each transition, while reducing 
the value of the cosmological constant, is almost always\footnote{The reason for the ``almost'' clause can be
understood from (\ref{viscs}) when, e.g. $p=3$, since the Cherns-Simons invariants are defined up
to an integer. } accompanied by a gauge symmetry breaking in some sector
due to the change in the discrete Wilson line data\footnote{Integers such as $w$ and $l$ which appear 
inside the threshold corrections (\ref{t3}) and (\ref{t4}) should also be regarded as
discrete Wilson line data, although their variation does not break any gauge symmetries.}. This mechanism strongly suggests that
the visible sector GUT symmetry breaking by the corresponding discrete Wilson line may have been triggered 
by a spontaneous nucleation of an M2-brane, leading to a simultaneous reduction of the cosmological constant.

When the visible sector three-cycle ${\cal Q}_{\rm vis}$ has a non-trivial fundamental group 
$\pi_1({\cal Q}_{\rm vis})={\bf Z}_p$, a discrete Wilson line of the form
\be
U={\rm diag}\left(e^{2\pi ik_1/p}\,,\,.\,.\,.\,,e^{2\pi ik_5/p}\right)\,,
\ee
where $\sum k_i=0$, is typically used to break the SU(5) GUT group. The corresponding Chern-Simons invariant
is given by \cite{Witten:1985mj}
\be
c_1^{\rm vis}=\sum_{i=1}^5\frac {k_i^2}{2p}\,\,\,\,\,\,{\rm mod\,\,{\bf Z}}\,.
\ee
When the gauge group is unbroken, $k_i=0\,\,\forall i$ and $c_1^{\rm vis}=0$.
To break $SU(5)\rightarrow SU(3)\times SU(2)\times U(1)$ we can choose the following discrete Wilson line 
data: $k_1=k_2=k_3=2w$ and $k_4=k_5=-3w$, where for $w$ we may pick one of the values $w=1,...,p-1$ such 
that $5w$ is not divisible by $p$. In this case, the fractional Chern-Simons invariant is 
\be\label{viscs}
c_1^{\rm vis}=\frac{15w^2}p\,\,\,\,\,\,{\rm mod\,\,{\bf Z}}\,.
\ee
Thus, apart from a couple of exceptional cases, the GUT symmetry breaking is automatically accompanied by a large
shift in the vacuum energy due to the spontaneous membrane nucleation. It is therefore not surprising 
that the energy scale obtained in the estimate (\ref{trans})
is of the same order of magnitude as the scale at which the visible sector SU(5) or SO(10) 
GUT gauge group is thought to be spontaneously broken to the Standard Model subgroup.

\acknowledgments I would like to thank Jacob Bourjaily, Jacques Distler, Michael Douglas, 
Katherine Freese, Ben Freivogel, Shamit Kachru, Stuart Raby, Gary Shiu,
Washington Taylor, Timo Weigand and Alexander Westphal for useful discussions and particularly acknowledge the input 
of my collaborators Bobby Acharya, Gordy Kane, Piyush Kumar, Jing Shao and Scott Watson without whom this work would not
have been possible. I would like to thank my wife Heather for support and SLAC for their hospitality.

\end{document}